# Magnetotransport properties of granular oxide-segregated CoPtCr films for applications in future magnetic memory technology

Morgan Williamson[1,2], Maxim Tsoi[1,2], Pin-Wei Huang, Ganping Ju, Cheng Wang [3]

[1]*Physics Department, University of Texas at Austin, Austin, Texas 78712, USA*

[2]*Texas Materials Institute, University of Texas at Austin, Austin, Texas 78712, USA*

[3]*Fremont Research Center, Seagate Technology, Fremont, California 94538, USA*

## ABSTRACT

Magnetotransport properties of granular oxide-segregated CoPtCr films were studied on both macroscopic and microscopic length scales by performing bulk and point-contact magnetoresistance measurements, respectively. Such a perpendicular magnetic medium is used in state-of-the-art hard disc drives and if combined with magnetoresistive phenomena (for read/write operations) may lead to a novel concept for magnetic recording with high areal density. While the bulk measurements on the films showed only small variations in dc resistance as a function of applied magnetic field (magnetoresistance of less than 0.02 %), the point-contact measurements revealed giant-magnetoresistance-like changes in resistance with up to 50,000 % ratios. The observed magnetorestive effect could be attributed to a tunnel magnetoresistance between CoPtCr grains with different coercivity. The tunneling picture of electronic transport in our granular medium was confirmed by the observation of tunneling-like current-voltage characteristics and bias dependence of magnetoresistance; both the point-contact resistance and magnetoresistance were found to decrease with the applied dc bias.

Magnetic data storage has been pushing scientific innovations and technology limits for more than half a century and has grown into a multi-billion-dollar industry today [1]. The recording of information in a magnetic medium, e.g. a thin magnetic film in a hard disk drive (HDD), is achieved by switching its magnetization locally between two different orientations, which represent "1" and "0" of digitally stored information. The way of storing bit information in granular medium with high magnetic anisotropy (Ku) is by far the most cost-effective technology for high density data storage, and has a clear road map of extendibility up to 2 TB/in$^2$. Such granular systems with high scalability and desired thermal stability can also be potentially leveraged into fabricating memory cells of non-volatile magnetic memory such as magnetic random access memory (MRAM). We thus envision a merging of hard drive granular medium and the MRAM concepts into a novel magnetic memory where reading and writing operations will be achieved by magnetoresistive transport means as in MRAM, while the thermally stable perpendicularly magnetized grains with diameters of below 10 nm provide extraordinary scaling potential for memory applications.

In this letter we explore magnetoresistive effects in the state-of-the-art perpendicular magnetic medium of today's HDDs – granular oxide-segregated CoPtCr films. We use point contact technique [2] to characterize the local transport in the film composed of granular layers with different coercivity. When the relative orientation of layers changes under the influence of an externally applied magnetic field, we observe large (up to 50,000 %) variations in the point-contact resistance. This magnetoresistance was found to decrease with the value of dc bias applied to the contact. Such variations are consistent with a tunnel magnetoresistance between individual CoPtCr grains and support the potential of the granular medium for future magnetic memory technology.

In our experiments we have tested a number of thin-film samples with various layer combinations, compositions, and individual layer thicknesses ranging from 0 to 100 Å. In what follows we focus mainly on the magnetoresistive behavior of a particular multilayer/sample: Ru(180)/M1(100)/Spacer(15)/M2(100)/Spacer(15)/M3a(100)/M3b(10) where all numbers in brackets are thickness in Å); with 63 nm of thick adhesion layer and seedlayer underneath, and a

6 nm thick continuous CoPtX magnetic capping layer (X is combination of non-magnetic diluting elements such as Cr, B, and Ru). M1, M2, M3a, and M3b are perpendicularly magnetized granular-media layers with different coercivity and a columnar structure of the metallic grain cores. The columnar structure is achieved by an epitaxy-like sputtering deposition [3] of the CoPtCr alloy [4] on top of a Ru-rich columnar growth template. Spacer layers are Ru-rich materials with oxide contents and are inserted to break the interlayer exchange coupling between adjacent ferromagnetic layers. Individual magnetic grains are segregated by a mixture of $TiO_2$, $SiO_2$ and other oxides. Figure 1(a) shows a cross-section transmission electron microscopy (TEM) image of the granular composite media. Figure 1(b) shows the magneto-optical Kerr effect (MOKE) hysteresis loop of a lone layer of the granular medium. The Kerr signal confirms the perpendicular anisotropy of the material and displays the effects of the magnetic grain switching field distribution and demagnetizing field as evidenced by the shearing of the major hysteresis loop.

Bulk magnetoresistive properties of the films were assessed by measuring the film resistance in the Van der Pauw geometry: four-point probes (wire-bonds) were placed in the corners of a 10×10 mm² square-shaped media film. An external magnetic field could be applied perpendicular to the film's plane (FPP-orientation) or in the plane of the film (FIP-orientation). Point contacts were then made to the thin-film samples and used to probe magnetotransport properties of small sample volumes ($10^6$-$10^9$ nm³) typically associated with such contacts. The point contact was made by bringing a sharpened metallic (Cu or W) tip into contact with the multilayer using a differential screw mechanism described elsewhere [2]. This mechanical point-contact technique enables us to produce electrical contacts of only a few nanometers in diameter (*a*) and probe electrical transport in very small sample volumes ~$a^3$. In this work we focus on dc transport measurements of point contacts in magnetic fields *B* up to 0.7 T applied in the FPP-orientation and dc biases *I* up to 1.5 mA. All measurements were done at room temperature.

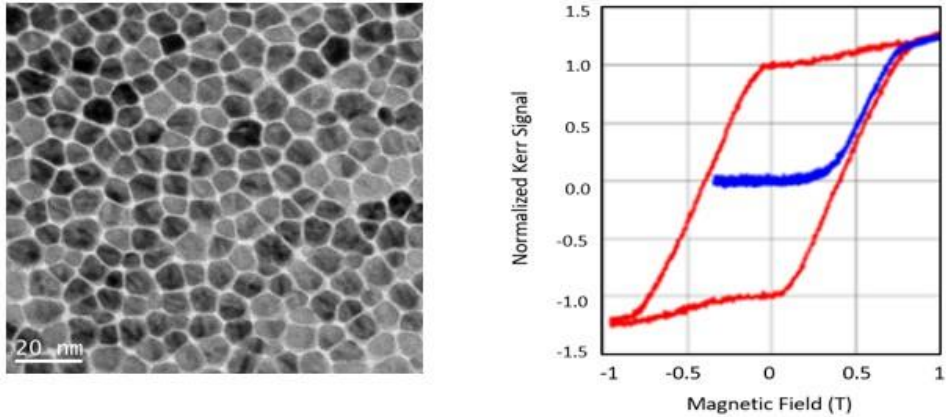

**Fig. 1** (a) Cross-section TEM image of the granular composite media. (b) Magnetic polar-Kerr loop of a granular-only sample. Note that the external field is slightly tilted from perpendicular to the sample surface with the purpose of magnifying the thermal decay effect (see the slope from saturation to zero-field remnant point).

We began by investigating the bulk transport properties of our samples. All samples showed a very small magnetoresistance (MR) with MR ratio ($\Delta R/R$) of about 0.02 % in the FIP geometry where the current flows primarily along the magnetic field direction. No MR was detected in the FPP orientation, and a continuous transition (decrease) in MR behavior was found between FIP and FPP orientations.

Contrary to the bulk measurements, point-contact measurements with Cu tips revealed a larger (~1%) magnetoresistance (MR) in both the FPP and FIP orientations. This still rather low MR ratio could originate from the relative softness of the copper compared to the hard media sample material. In order to make stable electrical contact with the sample, the copper point contacts required a high amount of pressure that visibly deformed the sharpened tips resulting in large contact areas consequently involving many individual grains in the sample. Due to the complications associated with the relatively soft copper tips, we have switched to tungsten as a point-contact tip material.

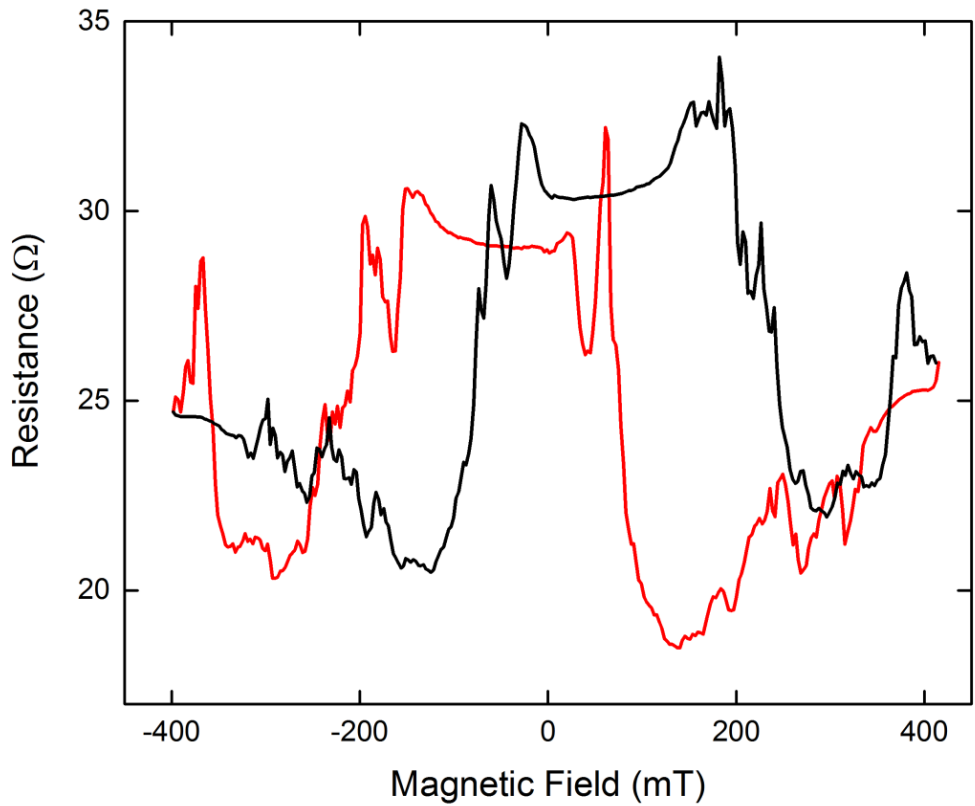

**Fig. 2** Point-contact magnetoresistance at 0.7 mA bias for a ±400 mT magnetic-field sweep displaying 50% MR ratio. Red (black) curve depicts magnetic-field downsweep (upsweep).

Figure 2 shows the resistance of a tungsten point contact under a 0.7 mA bias with an applied magnetic field in the FPP orientation. The measurement starts at a field with a magnitude just over +400 mT (red curve) pointing away from the sample surface. As the magnitude of the applied field is gradually swept to a value approaching 140 mT the resistance exhibits a low value of ~20 Ω. The resistance then rises as the device encounters 60 mT to a ~30 Ω demonstrating a 50% change in resistance. Interestingly, the device shows a relatively calm region from ~ 0 to -140 mT, a feature which becomes a signature of many of the tungsten point-contact measurements. Continuing the sweep to applied fields in the opposite direction, at ~ -300 mT the device resistance decreases to the low value of ~20 Ω. And after some vacillations, the resistance then increases slightly at the end of the downsweep to about 25 Ω. The upsweep (black curve) shows marked symmetry to the downsweep replicating both the calm plateau region at modest fields and the resistance spike just below ±400 mT. The hysteretic MR behavior displays symmetry about zero field, which with a properly tuned exchange bias field of ±150 mT could potentially be utilized as 50% MR ratio field-toggled device.

Along with the major reproducible features, like the calm plateau region, etc., MR traces show a fine structure which may vary from sweep to sweep. We illustrate the reproducibility and variability of MR traces in Fig. 3 which shows the major MR loop (bottom black and red curves) and various minor MR loops (starting at an applied field value slightly above +400 mT and having different turning point fields). Each subsequent measurement was shifted vertically to visually separate the trace from its neighbors. Five measurement repetitions are shown for each loop in light red (gray) corresponding to the magnetic downsweeps (upsweeps). The first minor loop doubles back at a field of -260 mT while each subsequent minor loop possesses a turning point field at -130, -60, 0, 70, and 100 mT, where red (black) curves represent a decreasing (increasing) magnetic field. The turning point of the first minor loop also coincides roughly with the resistance crossover between high and low resistance states in the major loop, effectively capturing the 50% MR behavior at the smallest negative value magnetic field. The second minor loop with its turning point of -130 mT does not access the low resistance state associated with high negative magnetic fields. Furthermore, the second, third, and fourth minor loops all show the similar behavior of remaining in the high resistance state near their turning points. The fifth minor loop with its turning point at 70 mT shows a hysteretic opening with slightly less area than the previous three minor loops, while the last minor loop fails to produce an appreciable hysteretic opening and remains in the low resistance state.

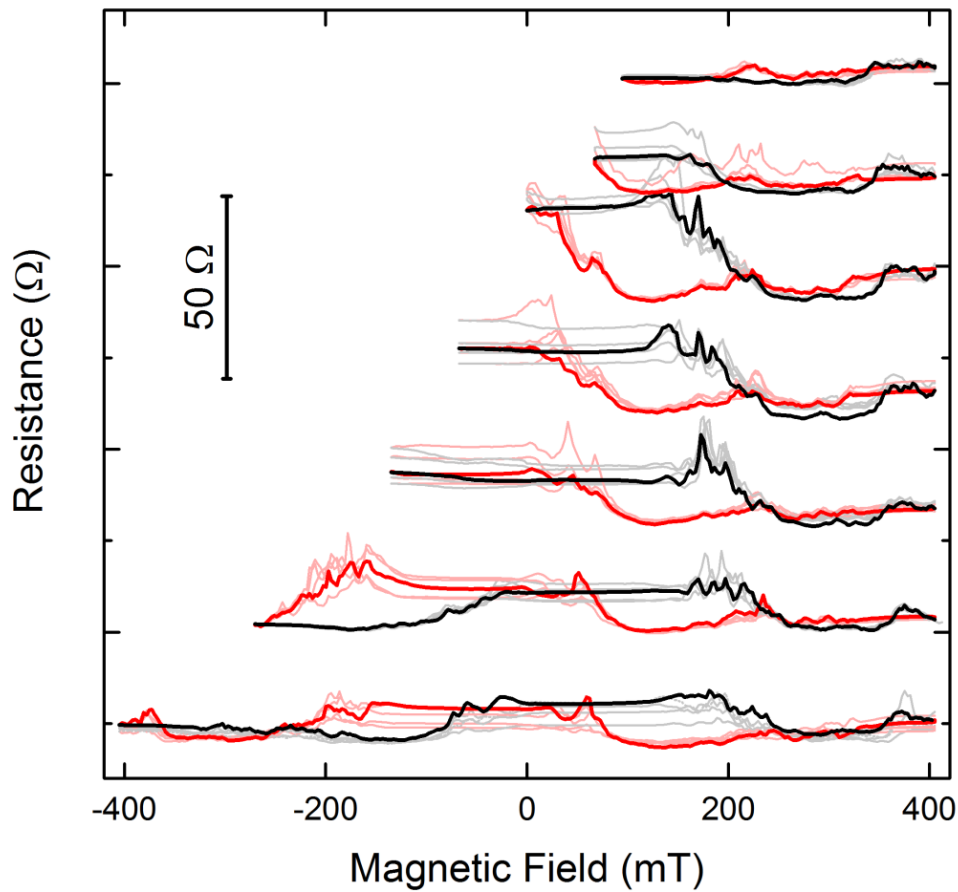

**Fig. 3** Major and minor MR loops showing magnetic hysteresis of the tungsten point contact at 0.7 mA bias. Different curves were shifted vertically for clarity. Each measurement starts at a field just over +400 mT. Red (black) curve depicts magnetic field downsweeps (upsweeps). Turning point fields are -400, -260, -130, -60, 0, 70, and 100 mT. Lighter colored red/gray curves indicate downsweep/upsweep repetitions of the same measurement while darker colored curves highlight representative sweeps.

Figure 4 shows an example of point-contact magnetoresistance with an MR ratio of >10,000%. Compared with the contact from Fig. 2 the principal modification is the resistance of the high resistance state ($R_H$), while the low resistance state ($R_L$) and the general form of the

MR curve is retained. At high fields, and presumably saturated coercivity among all layers, the resistance $R_L$ is measured at 20 Ω, while at modest fields there are large variations in resistance hovering between 2 kΩ and 6 kΩ. We observed a strong dependence of the MR on the bias applied to the contact. Lighter colored curves in Fig. 4 show MR of the same device under a different (higher) bias current of 1.5 mA that results in a 1500 % MR ratio. The insert to Fig. 4 shows that the MR ratio decreases with increasing bias independent of the bias polarity. We note that the main change in the magnitude of MR originates in the bias dependence of $R_H$ while $R_L$ remains relatively intact.

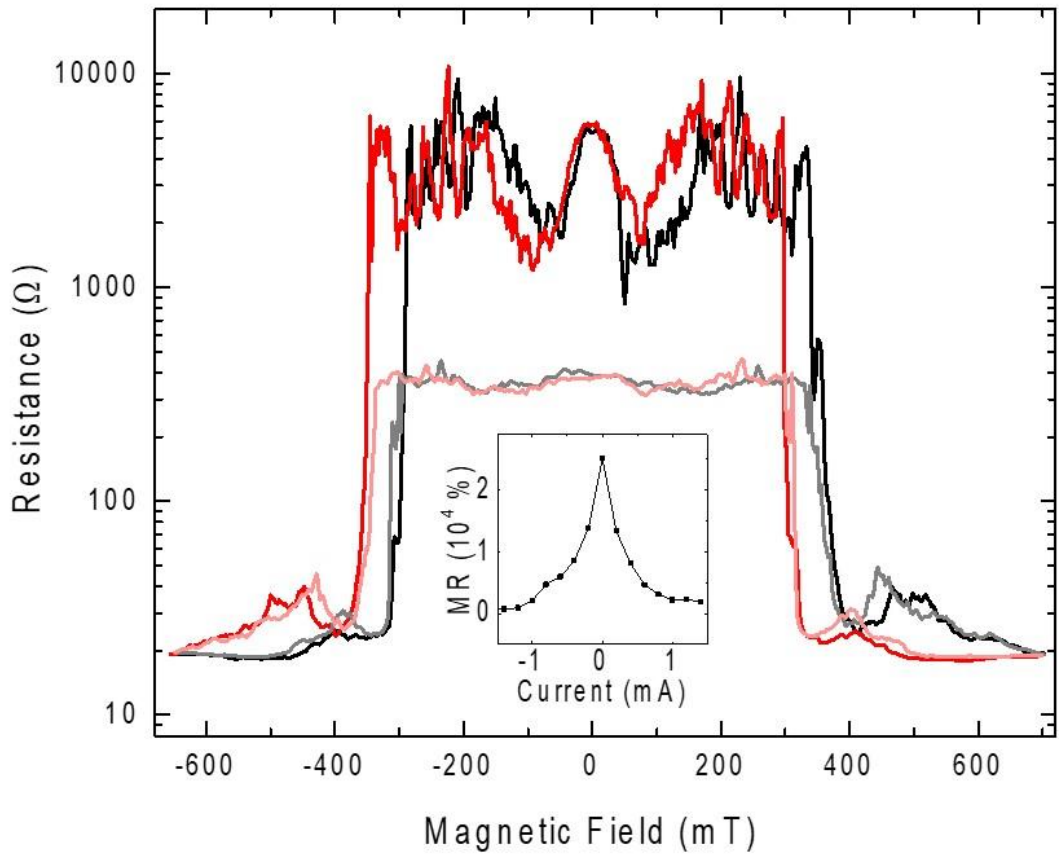

**Fig. 4** Magnetoresistance trace of tungsten PC device displaying vast variation in resistance from 20 Ω to 4 kΩ corresponding to >10,000% MR. Red (black) curve depicts magnetic field downsweep (upsweep) recorded at low applied bias of 0.1 mA. Lighter

colored curves show MR of the same device under a different (higher) bias current of 1.5 mA (1500 % MR ratio). The insert shows the bias dependence of MR ratio.

A point contact between two metallic electrodes at room temperature usually assumes a diffusive metallic conduction [5]. The latter cannot explain the very high resistance (almost 10 kΩ) of our point contact. We speculate that the large resistance changes demonstrated in Fig. 4 may be caused by tunneling effects, which we would expect to be relevant in our composite sample material. The Co-Pt-Cr granular media under study has a hexagonal columnar structure that is perpendicular to the plane and would presumably support high current densities in the perpendicular direction. The perpendicularly magnetized columns, separated by oxide barriers, would serve as channels for these perpendicular currents. However, an adjacent layer comprised of similar magnetic columnar grains with opposite magnetization may hinder perpendicular current flow, perhaps bringing it to a point on par with the small in-plane currents i.e. between channels across the oxide segregation barriers. If perpendicular currents and in-plane currents have similar path resistances the total perpendicular current will be funneled to the relatively few magnetic domain interfaces that exhibit aligned orientations. This process, active under non-saturating applied fields, would effectively create a network of tunnel junctions composed of both intralayer and interlayer tunnel junctions leading to possibly very large resistance changes. Observing such behavior would require a probe with a similar contact area as the magnetic grain size, namely, 7–9 nm diameter since probes with larger surface areas would average over many grains overlooking the nanoscale behavior. Achieving a nanoscale probe area motivated the introduction of a tungsten point-contact tip as discussed. Tungsten, due to its hardness, is expected to deform less when contacting the sample surface. Tungsten point contacts would therefore possess a smaller contact area more suitable for investigating the nanoscale behavior of the granular composite media and associated large resistance changes caused by the proposed tunnel junction network.

In conclusion, we investigated the bulk and point-contact (local) transport properties of CoPtCr granular composite media films. While the bulk measurements on the films showed only small variations in dc resistance as a function of applied magnetic field (magnetoresistance of less than 0.02 %), the point-contact measurements revealed giant-magnetoresistance-like changes in resistance with up to 50,000 % ratios. The observed magnetorestive effect is tentatively attributed to a tunnel magnetoresistance between CoPtCr grains with different coercivity. The tunneling picture of electronic transport in our granular medium was confirmed by the observation of tunneling-like current-voltage characteristics and bias dependence of magnetoresistance; both the point-contact resistance and magnetoresistance were found to decrease with the applied dc bias. The observed magnetotransport effects support the feasibility of a novel concept of magnetic memory based on granular composites where reading and writing operations would be achieved by magnetoresistive transport means.

This work was supported in part by the Sponsored Research Agreement No. UTA18-000691 between the University of Texas at Austin and Seagate Technology.